\documentclass[a4paper]{article}

\usepackage[margin=2cm]{geometry}
\usepackage[utf8]{inputenc}
\usepackage{bm}
\usepackage{graphicx}
\usepackage{gensymb}
\usepackage{url}
\usepackage{hyperref}
\usepackage{braket}
\usepackage{textcomp}
\usepackage{multirow}
\usepackage{subcaption}
\usepackage{color}
\usepackage{pdfpages}
\usepackage{amssymb}
\usepackage{amsmath}
\usepackage{comment}
\usepackage[dvipsnames]{xcolor}
\usepackage[normalem]{ulem}
\usepackage{hyperref}
\usepackage{natbib}

\newcommand{\gao}[0]{Ga$_2$O$_3$}
\newcommand{\bgao}[0]{$\beta$-Ga$_2$O$_3$}
\newcommand{\ggao}[0]{$\gamma$-Ga$_2$O$_3$}

\newcommand{\fluence}[2]{$#1\times10^{#2}$ cm$^{-2}$}
\newcommand{\num}[2]{$#1\times10^{#2}$}

\begin{document}

\title{From phase transformation to amorphization: damage accumulation in Yb-implanted \bgao} 
\date{}
\maketitle

\author{Joanna Matulewicz$^{1,2*}$, 
Renata Ratajczak$^{1}$, 
Ewa Grzanka$^{3}$, 
Maciej Oskar Liedke${^4}$,
Damian Kalita$^{1}$, 
Eric Hirschmann${^4}$,
Andreas Wagner${^4}$,
Mikołaj Grabowski${^3}$,
Michał A. Stróżyk${^1}$,
Cyprian Mieszczyński$^{1}$,
Przemyslaw Jóźwik$^{1}$, 
Ulrich Kentsch$^{4}$, 
Rene Heller$^{4}$,
Frederico Garrido${^2}$}     \\

${^1}$ National Centre for Nuclear Research, Andrzeja Sołtana 7, 05400 Otwock, Poland

${^2}$ Université Paris-Saclay, CNRS/IN2P3, IJCLab, 91405 Orsay, France

${^3}$ Institute of High Pressure Physics UNIPRESS, Polish Academy of Sciences, Sokolowska 29/37, 01142 Warsaw, Poland

${^4}$ Helmholtz-Zentrum Dresden-Rossendorf, Bautzner Landstrasse 400, 01328 Dresden, Germany \\


${^*}$ corresponding author: joanna.matulewicz@ncbj.gov.pl

%% Abstract
\begin{abstract}

This study provides a comprehensive analysis of the radiation response and structural evolution of differently oriented \bgao\ single crystals, subjected to Yb ion implantation over a wide fluence range from \num{5}{12} to \fluence{1}{16} (0.04–74 dpa). A multi-technique approach (RBS/C, PAS, HRTEM, and HRXRD) was employed to investigate the mechanisms of damage accumulation. 
The results reveal a multi-stage process of defect evolution. At a critical threshold of around 0.4 dpa, the accumulation of lattice strain triggers a phase transformation from monoclinic \bgao\ to a defective spinel structure of \ggao. Notably, the formation of this new phase is accompanied with strain relaxation. With further irradiation, defects develop within the crystal structure of \ggao. The associated atomic reorganization at this stage is reflected by a distinct dip in the damage accumulation curve and the appearance of stacking faults in the subsurface region of the implanted layer.
In contrast to previous reports suggesting high radiation stability of this phase, the present study clearly demonstrates that continuous defect accumulation results in a significant increase in both displaced atoms and vacancy-type defects, with a strong depth dependence in their type and density. Ultimately, at an irradiation level of approximately 7 dpa, amorphization of the surface layer occurs. With further irradiation, the amorphous layer thickens, gradually replacing the $\gamma$ phase.
These findings reveal that the radiation tolerance of gallium oxide is highly sensitive to ion-specific interactions and strain-induced instabilities, thereby challenging the previously assumed robustness of this material under high-fluence ion irradiation.

\end{abstract}

\section{Introduction}
\label{intro}

One of the most promising materials that is extensively studied nowadays is gallium oxide (\gao). With a bandgap in the range of 4.5-5.7 eV, \cite{ganguly2025advances} \gao\ belongs to the ultra-wide bandgap semiconductors (UWBG). UWBG possess several important qualities, such as a high breakdown field and resistance to high temperature and voltage \cite{xu2022review}, which position \gao\ as a key material for next-generation power electronics, optoelectronics, and scintillation applications \cite{galazka2018beta}. \gao\ is also considered to be radiation resistant \cite{azarov2023universal}. These characteristics might allow the \gao-based devices to work in harsh conditions, such as space systems or nuclear reactors \cite{titov2022comparative}.

Gallium oxide exhibits a rich polymorphism, as it can crystallize in several phases: $\alpha$, $\beta$, $\gamma$, $\delta$ , $\epsilon$ and $\kappa$  \cite{xu2022review}. Out of them, the monoclinic $\beta$ phase is the most thermodynamically stable under normal conditions \cite{pearton2025perspective}, maintaining the crystalline structure up to its melting point of 1780 \textdegree C \cite{ganguly2025advances}. The crystal structure of \bgao\ is complex. Its monoclinic unit cell contains two inequivalent Ga sites (tetrahedral and octahedral) and three inequivalent O sites. Consequently, the material exhibits pronounced anisotropic properties. As demonstrated by studies of various crystal orientations, this anisotropy notably affects thermal conductivity \cite{slomski2017anisotropic}, optical luminescence \cite{matulewicz2025comprehensive}, and absorption \cite{ricci2016theoretical}. 

The fundamental properties of \bgao\ can be further modified through doping. In particular, doping with rare-earth (RE) ions enables precise tuning of the optical properties \cite{zhang2025rare}. 
Because of the wide bandgap of the host material, the thermal quenching of luminescence efficiency from RE ions is significantly reduced \cite{favennec1989luminescence}, enhancing the optical performance of the \gao:RE systems. To overcome the inherently low solubility limits of RE, ion implantation is often employed as an effective doping method  \cite{majid2012neon}. This technique allows for precise control of the dopant concentration and its depth distribution profile by beam energy and ion fluence. Nevertheless, the process of ion implantation causes damage in the crystal lattice and/or phase transformations, which can alter the material's performance. 

Simultaneously, ion implantation can be used as an effective method for evaluating radiation resistance, defined as the ability of the crystal structure to remain preserved under irradiation. It has been widely documented that \bgao\ undergoes a phase transformation to \ggao\ when exposed to radiation \cite{garcia2022formation,zhao2025crystallization}. This phenomenon is currently a subject of wide research.
The defective spinel cubic \ggao\ is the least stable form of gallium oxide \cite{sun2025recent}, and typically transforms to \bgao\ at 490-575\textdegree C \cite{tang2024thermal}. 
Several reports show that the formed $\gamma$ phase does not undergo amorphization even at very high ion fluences, reaching 265 dpa \cite{azarov2023universal,wendler2016high,azarov2022disorder,bektas2025defect}, suggesting its high radiation resistance.
In contrast, other studies have reported amorphization occurring at around 7 dpa following ytterbium (Yb) implantation  \cite{ratajczak2024anisotropy}. 
Similar suggestions based on RBS/C spectra after europium (Eu) implantation at a comparable dpa level \cite{lorenz2014doping,peres2017doping}, as well as the disappearance of the \ggao\ XRD signal after boron (B) implantation with a fluence corresponding to approximately 7 dpa \cite{nikolskaya2025structure}, further support this amorphization threshold. 
Consequently, radiation resistance remains a complex issue, potentially depending on the physicochemical properties of the implanted ions and/or the implantation conditions.

In this study, (010)- and ($\bar{2}$01)-oriented \bgao\ crystals were implanted with Yb ions with a wide range of fluences, from \num{5}{12} up to \fluence{1}{16}. The work aims to provide a comprehensive understanding of the damage accumulation process in \bgao:Yb by examining each stage with multiple experimental methods. Our previous study reported by \citet{sarwar2024defect} revealed the general course of the process, indicating additional structural disturbances at higher ion implantation fluences above the threshold for \ggao\ formation. One noticeable anomaly was observed at a Yb fluence of approximately \fluence{4}{14}, which was manifested as a drop in the RBS/C-derived damage accumulation curve. Furthermore, the studies suggested pronounced lattice deformation at around \fluence{1}{15}. Consequently, in the present work, special attention is paid to these specific fluence regimes to elucidate the underlying structural changes.

The implanted crystals were examined with several complementary experimental methods. Rutherford Backscattering Spectrometry in channeling mode (RBS/C) was used to establish the damage accumulation curve and identify potential orientation-dependent differences. RBS/C is a powerful method in studying the crystal quality because it provides information on the depth distribution of defects. However, it has significant diagnostic limitations. Specifically, it is hardly sensitive to vacancy-type defects, thereby preventing the unambiguous identification of the defect types or precise determination of the physical mechanisms behind the observed phase and structural transformations. For this reason, High Resolution X-ray Diffraction (HRXRD) was employed to evaluate the radiation-induced phase changes and lattice strain in the material. Furthermore, High Resolution Transmission Electron Microscopy (HRTEM) was used to examine and identify the post-implantation layers. In addition, Positron Annihilation Spectroscopy (PAS) was applied to analyze the size and concentration of vacancies and vacancy-related clusters.

\section{Experimental}
\label{exp}

\subsection{Sample preparation}
\label{sp}

The studies were performed on commercially available unintentionally doped (UID) \bgao\ crystal substrates with (010) and ($\bar{2}$01) orientations, manufactured by Tamura Corporation. The \bgao\ substrates, sold as 2 in. wafers with a thickness of 650 $\mu$m, were diced into smaller samples of approximately 9$\times$9 mm. To protect the surface during the dicing process, a protective layer was deposited, which was removed by organic cleaning afterwards.

Ion implantation was carried out at room temperature using 150 keV Yb ions with fluences ranging from \num{5}{12} to \fluence{1}{16}, which corresponds to a 0.04-74 dpa range. To avoid any channeling effect during implantation, the samples were tilted 7\textdegree\ relative to the ion beam direction. The ion implantation process was conducted in the Ion Beam Center (IBC) at Helmholtz Zentrum Dresden-Rossendorf (HZDR), Dresden, Germany.

\subsection{Analytical methods}
\label{AM}

RBS/C measurements were carried out at IBC, HZDR, using a 2 MV Van de Graaff accelerator with a 1.7 MeV $^4$He$^+$ ion beam. The measurements were performed in a chamber equipped with a three-axis goniometer and a silicon detector, positioned at a 170\textdegree\ scattering angle. This setup provided a depth resolution below 5 nm and an energy resolution below 20 keV.

PAS experiments were performed at the radiation source ELBE, HZDR. Doppler Broadening - Variable Energy Positron Annihilation Spectroscopy (DB-VEPAS) measurements were performed using the Apparatus for In-situ Defect Analysis (AIDA) \cite{liedke2015open} integrated into the slow-positron beamline SPONSOR \cite{anwand2012design}. The kinetic energy of implanted positrons was in the range 0.05-35~keV, enabling depth profiling. More details can be found in the supplementary materials. 
Variable Energy Positron Annihilation Lifetime Spectroscopy (VEPALS) measurements were carried out at the Mono-energetic Positron Source (MePS) beamline of the radiation source ELBE, at HZDR \cite{wagner2018positron}. The gamma radiation, resulting from annihilation events, was detected with a CeBr$_3$ scintillator detector coupled to a Hamamatsu R13089-100 photomultiplier tube, and the signals were processed by a Teledyne SP Devices ADQ14DC-2X digitizer \cite{hirschmann2021new}. The overall time resolution of the measurement system was better than 230 ps, and all spectra contained at least \num{1}{7} counts. More details can be found in supplementary materials.

HRTEM imaging was carried out in the National Centre for Nuclear Research (NCBJ), Otwock, Poland, using a JEOL JEM-F200 microscope operating at 200 kV. Thin samples were prepared with a focused ion beam (FIB) lift-out method using a ThermoFisher Scientific Helios 5 UX dual-beam scanning electron microscope (SEM). To avoid the formation of Ga$^+$-induced structural defects, the final polishing was performed with the acceleration voltage of 2 kV.

HRXRD scans were performed at the Institute of High Pressure Physics (UNIPRESS), Polish Academy of Sciences, Warsaw, Poland, using an Empyrean X-ray diffractometer operating at Cu K$\alpha_1$ wavelength (40~kV, 30~mA), equipped with a hybrid 2-bounce monochromator and a threefold Ge(220) analyzer. 2$\theta$/$\omega$ scan of symmetrical reflection was done for each sample.

\subsection{Computer simulations}
\label{sim}

SRIM simulations \cite{ziegler2010srim} were used to calculate the displacement per atom (dpa) values, assuming the default values of Ga and O threshold displacement energies ($E_d$) given by the SRIM software, 25 and 28 eV, respectively. The value was calculated according to the formula
\[
\mathrm{dpa} = \frac{T_{max}\times\Phi}{n_{at}},
\]
where $T_{max}$ represents the maximum of the atomic displacement distribution (including vacancies and atomic replacements from the VACANCY.txt and NOVAC.txt files), $\Phi$ is the ion fluence expressed as the number of ions per cm$^2$, and $n_{at}$ is the atomic density of \bgao, equal to \num{9.45}{22} cm$^{-3}$ (corresponding to 5.88 g/cm$^3$). 

The value of dpa obtained with the default $E_d$ for 150 keV Yb ions at a fluence of \fluence{1}{14} is equal to 0.74. For comparison, using alternative $E_d$ values of 23 eV for Ga and 17 eV for O \cite{he2024threshold} yields a higher value of 1.10 dpa. The significant difference in calculated dpa values (0.74 vs. 1.10) highlights the sensitivity of radiation damage estimates to selected displacement threshold energies $E_d$. To maintain consistency with previous results, default values of $E_d$ were used here.

\section{Results}
\label{results}

\subsection{Defect accumulation curve}

\begin{figure}[h]
    \centering
    \includegraphics[width=1\linewidth]{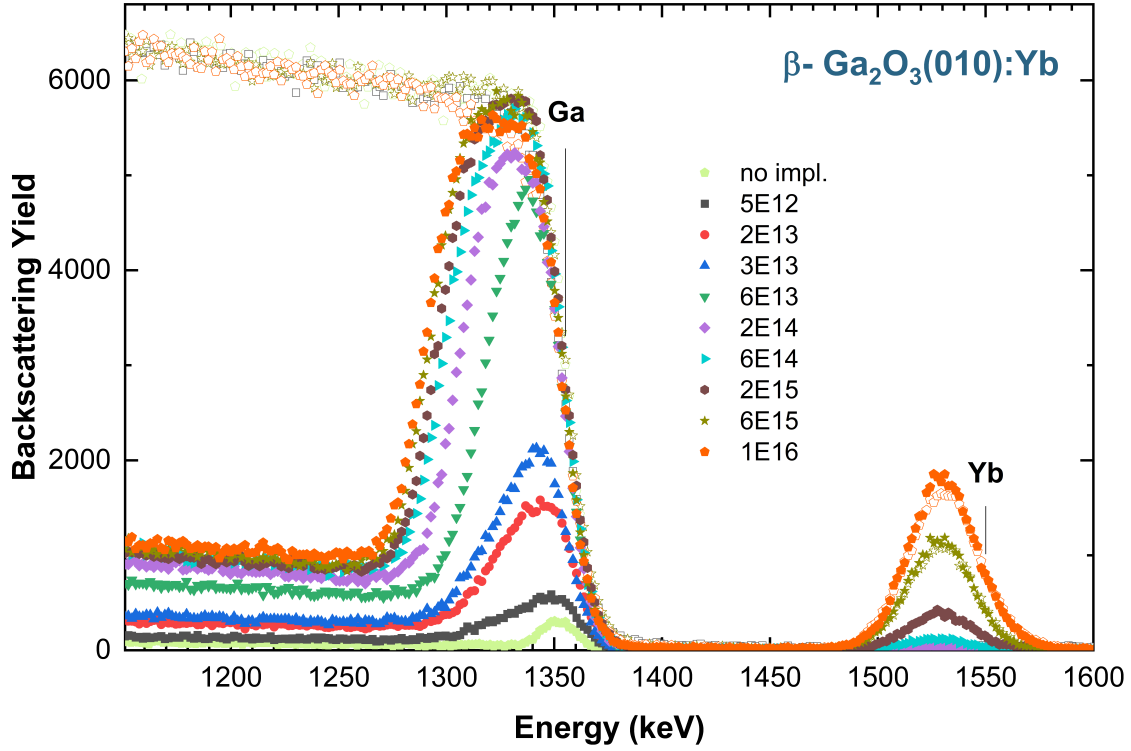}
    \caption{RBS/C spectra of (010)-oriented \bgao\ implanted with different Yb ion fluences, expressed in cm$^{-2}$. Solid symbols indicate spectra recorded in the axial direction, while empty symbols indicate corresponding spectra recorded in the random direction.}
    \label{fig:rbs_example}
\end{figure}

RBS/C is a powerful tool to study the crystal quality with depth scale. The number of backscattered He ions, including both direct scattering and dechanneling components, provides information about the concentration of single and extended defects.
Representative RBS/C spectra collected for RE-implanted \bgao\ are presented in Fig. \ref{fig:rbs_example}. 
Due to the much higher atomic mass of Yb compared to Ga and O, and because Yb atoms are incorporated only up to a limited depth, the signal originating from He ions backscattered from Yb appears as a distinct peak centered at approximately 1530 keV. The lower-energy part of the spectrum (below 1400 keV) corresponds to backscattering from Ga atoms. 
The signal originating from the Ga atoms is conventionally employed to establish the damage accumulation curve.
By analyzing the damage peak observed in the 1290-1355 keV region, we can track the evolution of the defect concentration in the host lattice as a function of ion fluence.

\begin{figure}[h!]
    \centering
    \includegraphics[width=0.85\linewidth]{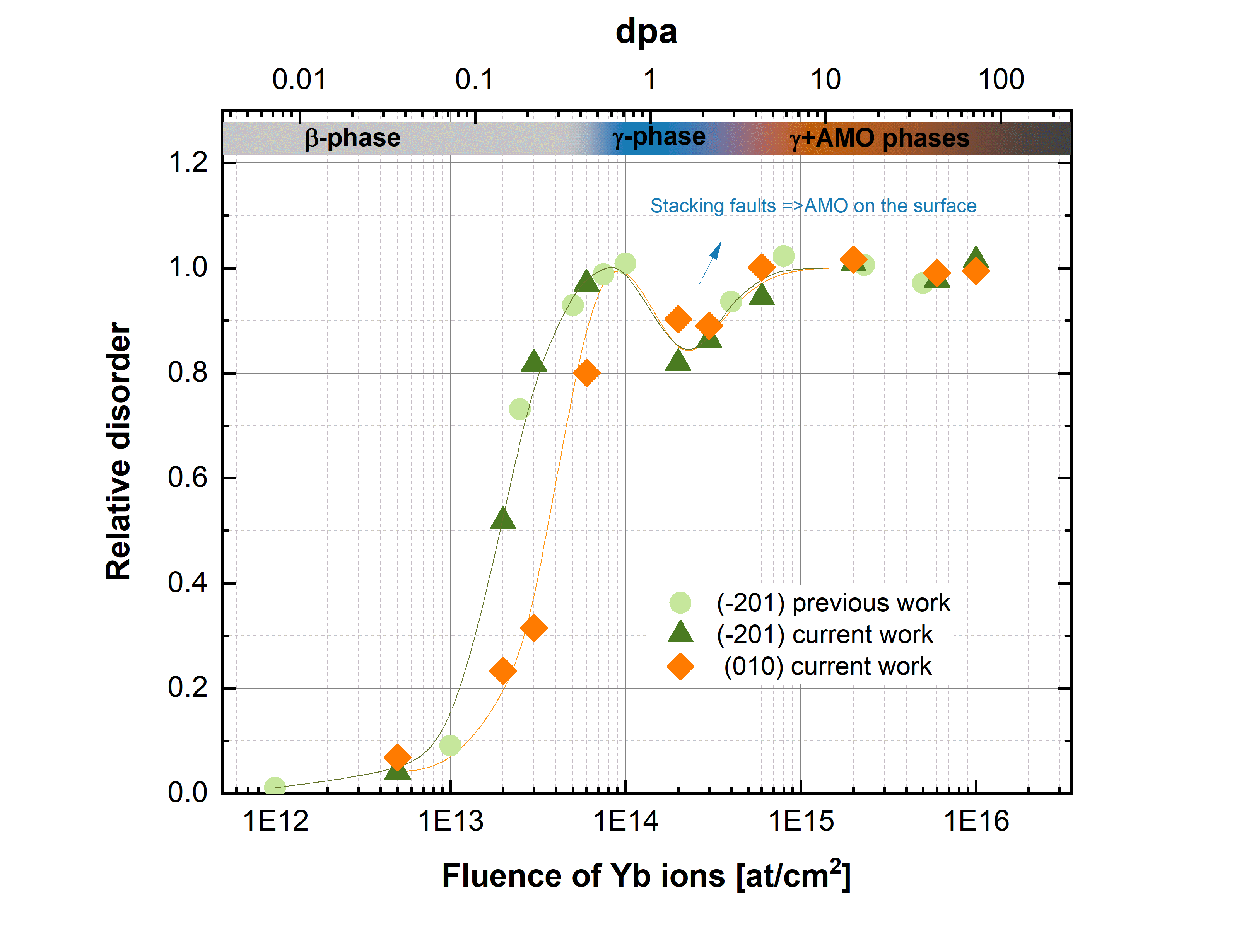}
    \caption{Damage accumulation curve for ($\bar{2}$01)- and (010)-oriented \bgao, showing the relative disorder as a function of Yb ion fluence. The bottom and top axes represent Yb ion fluence, expressed in ions/cm$^2$ and in dpa, respectively. Previous measurements for the ($\bar{2}$01)-oriented \bgao\ \cite{sarwar2024defect} are included as a reference to demonstrate data consistency and reproducibility. Lines are drawn to guide the eye and do not represent an analytical fit.}
    \label{fig:damage_accumulation}
\end{figure}

The relative disorder, determined by calculating the ratio of the aligned to random RBS signals coming from Ga atoms, is presented in Fig. \ref{fig:damage_accumulation}. 
This calculation was performed within a selected energy window (1320–1340 keV) located immediately beyond the surface peak. This region corresponds to the maximum of the damage peak and covers the near-surface region down to 30 nm. 
The values of relative disorder were calculated for both (010)- and ($\bar{2}$01)-oriented \bgao\ across an ion fluence range from \num{5}{12} to \fluence{1}{16}, and reflect the key steps of damage evolution observed in these crystals. 
Fig. \ref{fig:damage_accumulation} also presents results from our earlier work  on ($\bar{2}$01) \bgao:Yb \cite{sarwar2024defect}, demonstrating full agreement and complementarity between the two datasets. 
It should be added that the McChasy simulations \cite{mcchasy} performed in support of the past investigations  indicated that the damage profile becomes bimodal at a fluence of \fluence{4}{14}. 
This evolution is accompanied by a reduction in defect concentration within the subsurface layer, suggesting the formation of two distinct defects types localized in separate depth regions. 
Therefore, in this present work, this fluence region was studied in greater detail.

Following the previously established classification \cite{sarwar2024defect}, the damage accumulation process in Yb-implanted \bgao, presented in Fig. \ref{fig:damage_accumulation}, can be divided into four distinct fluence regions.
Region I, up to \fluence{1}{13}, is characterized by a slow increase in damage. 
In region II,  extending up to \fluence{1}{14}, a very rapid increase in the relative disorder is observed, corresponding to the $\beta$- to \ggao\ phase transformation \cite{sarwar2024defect}. 
The misorientation between the newly formed \ggao phase and the matrix enhances dechanneling, as the probing ion beam is aligned along the \bgao\ crystal axis. Consequently, the RBS aligned yield increases, even in the absence of amorphization. 
In region III, a reduction of the backscattering yield within the subsurface region is observed. It is attributed to structural reorganization that facilitates better alignment with the \bgao\ crystal axis, rather than to a decrease in defect concentration.
In region IV, above \fluence{1}{15}, the backscattering yield once again reaches saturation at the random level, which can be ascribed to the end of the second structural transformation process and the formation of a stable, highly disordered or modified crystalline state. 

The character of the damage accumulation kinetics observed for (010)-oriented \bgao\ is similar to that observed for ($\bar{2}$01)-oriented crystals. In both cases, the $\beta$ to $\gamma$ phase transition process starts at approximately 0.1 dpa and is completed at around 0.7 dpa. Since the transformation occurs at approximately the same fluences for (010)- and ($\bar{2}$01)-oriented crystals, it can be concluded that this process is independent of crystallographic orientation.

In contrast, MD simulations have suggested that the phase transition in \bgao\ is orientation-dependent, with the critical fluence required for the $\beta$ to $\gamma$ transformation in (010)-oriented \bgao\ being approximately 30\% higher than for ($\bar{2}$01)-oriented crystals \cite{liu2026anisotropic}. According to the authors, the ($\bar{2}$01) surface orientation is non-channeling, whereas the (010) orientation is strongly channeling. This is supported by simulations showing deep ion penetration for the (010) orientation, despite a 7\textdegree\ tilt angle, while for the ($\bar{2}$01) orientation, implantation-induced defects are confined to a relatively shallow depth \cite{liu2025orientation}.

Experimentally, neither a higher critical fluence for the $\beta$ to $\gamma$ phase transition (presented here), nor a deeper damage profile \cite{ratajczak2024anisotropy} is observed for the (010)-oriented samples compared to the ($\bar{2}$01)-oriented ones. However, this observation should not be interpreted as a contradiction of the MD predictions, since the expected orientation-dependent shift of the damage profile may be comparable to the experimental depth resolution of the presented RBS/C measurements (about 10 nm). 

Nevertheless, the relative disorder in region II for the (010) orientation is consistently lower than for the ($\bar{2}$01) orientation, typically by about 10\%. 
According to McChasy simulations, this discrepancy cannot be solely explained by the difference in the channeling cross-section between the (010) and ($\bar{2}$01) orientations of \bgao\ \cite{ratajczak2024anisotropy}. 
Therefore, the experimentally observed orientation dependence of the RBS/C response likely involves an additional orientation-dependent contribution, beyond conventional channeling effects. One possible explanation is that the implantation-induced defect structure may develop an anisotropic morphology, resulting in an orientation-dependent interaction with the probing He ion beam. We propose that the generated atomic displacements and the associated early-stage defect clusters may include configurations with orientation-dependent visibility. Due to the wide, open [010] crystallographic channels and geometrical shielding effects, such defect configurations may experience reduced interaction with the probing He ions. As a result, the probing He ions may undergo reduced dechanneling and consequently have a lower probability of direct backscattering, leading to a seemingly lower damage level in region II of the damage accumulation curve. In contrast, along the direction perpendicular to the ($\bar{2}$01) plane, the narrower and more restricted channels are expected to be more sensitive to the same defect configurations, resulting in enhanced dechanneling and direct backscattering. Thus, the orientation-dependent visibility of irradiation-induced defect structures provides a possible explanation for the seemingly higher relative disorder observed for the ($\bar{2}$01) orientation and the larger concentration of extended defects \cite{matulewicz2025comprehensive}, even though both orientations appear to undergo the phase transformation simultaneously. 
Another possible explanation is that the atomic reorganization during the nucleation of crystalline \ggao\ phase (specifically, the (110) orientation on (010)-oriented \bgao, and (111) on ($\bar{2}$01)-oriented \bgao\  \cite{bektas2025defect}) exhibits varying degrees of misorientation relative to the principal crystallographic axes of the host substrate. This structural misalignment may enhance ion dechanneling in RBS/c even before long-range order of the \ggao\ layer is detectable by XRD.

\subsection{Vacancy-related defect behavior}

\begin{figure}[h!]
        \centering
        \includegraphics[width=0.7\linewidth]{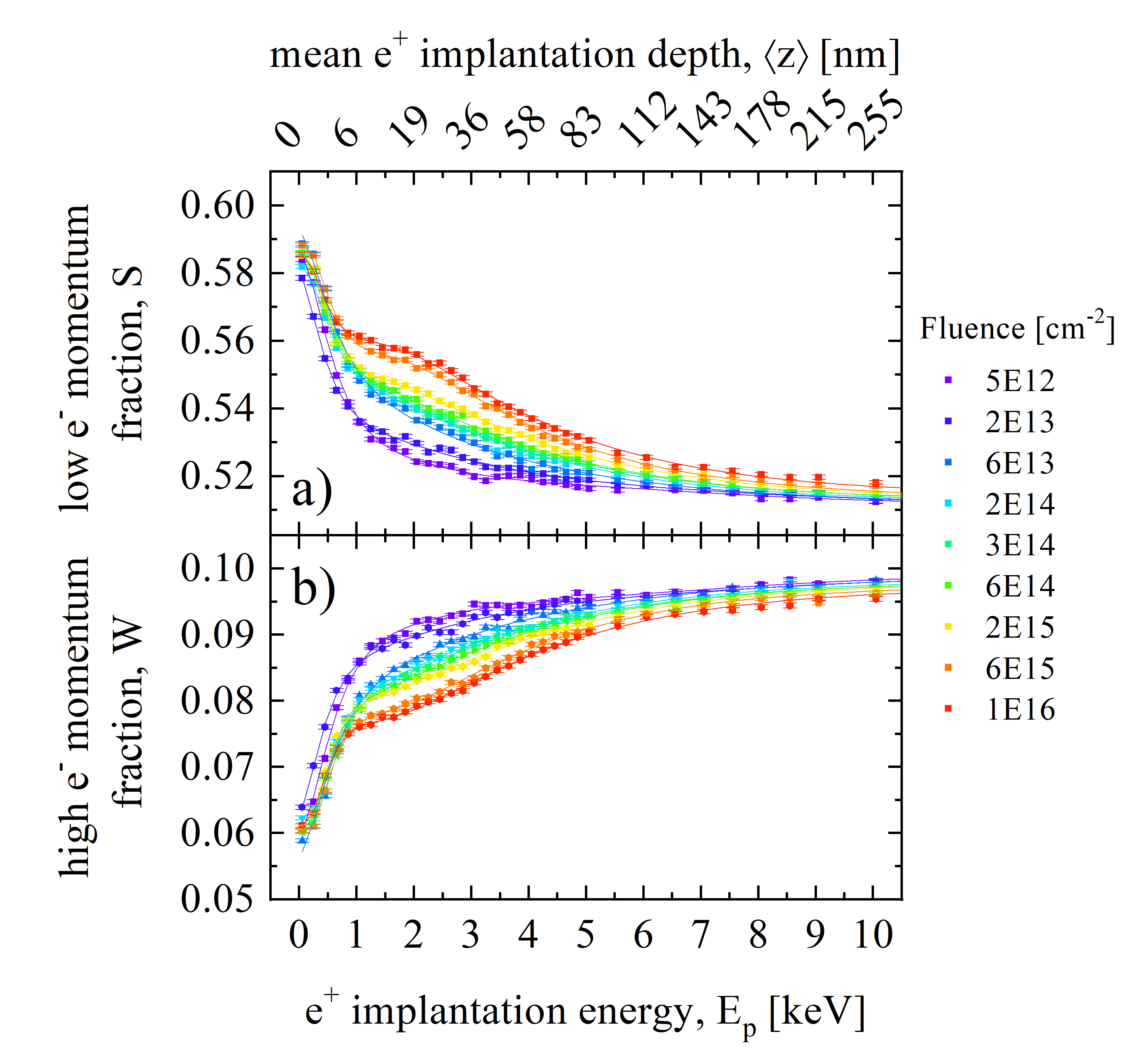}
        \caption{S- and W-parameter as a function of positron implantation energy and corresponding mean implantation depth, fitted using VEPfit code \cite{Veen1991,Veen1995}. The results are for ($\bar{2}$01)-oriented Yb-implanted \bgao\ crystals.}
        \label{fig:sw_energy}
\end{figure}

Positron annihilation spectroscopy (PAS) is based on the observation that the positron annihilation parameters, such as the positron lifetime and momentum distribution, are sensitive to the presence of open-volume defects, such as vacancies, vacancy agglomerates, and dislocations \cite{krause-rehberg2003positron}. By employing Doppler Broadening Variable Energy Positron Annihilation Spectroscopy (DB-VEPAS), it is possible to distinguish two types of positron annihilation events. Annihilations with valence electrons contribute to the low electron momentum fraction, called the S-parameter. In contrast, the W-parameter represents the fraction of positrons that annihilate with high-momentum core electrons. Fig. \ref{fig:sw_energy} shows the S- and W-parameters as a function of the positron energy $E_p$ and corresponding mean positron implantation depth for various Yb fluences in ($\bar{2}$01)-oriented \bgao\ crystals. The results were fitted using VEPfit code \cite{Veen1991,Veen1995}. Since the S- and W-parameters scale with defect density, the observed increase of S (and decrease of W) with Yb ion implantation fluence, seen in Fig. \ref{fig:sw_energy}, indicates increasing concentration and/or size of implantation-induced defects. This method is sensitive to neutral and negatively charged vacancies, therefore positively charged defects, such as oxygen vacancies, remain undetected due to the Coulomb repulsion of positrons. 
The observed step-like slope in Fig. \ref{fig:sw_energy} suggests an accumulation of defects in the near-surface region (mostly $<60$ nm), which reflects the limited diffusion of positrons from the mean depth given by the implantation energy and corresponds approximately to the thickness of the implanted layer \cite{ratajczak2024anisotropy,sarwar2024defect}. 

\begin{figure}[h!]
        \centering
         \includegraphics[width=0.5\linewidth]{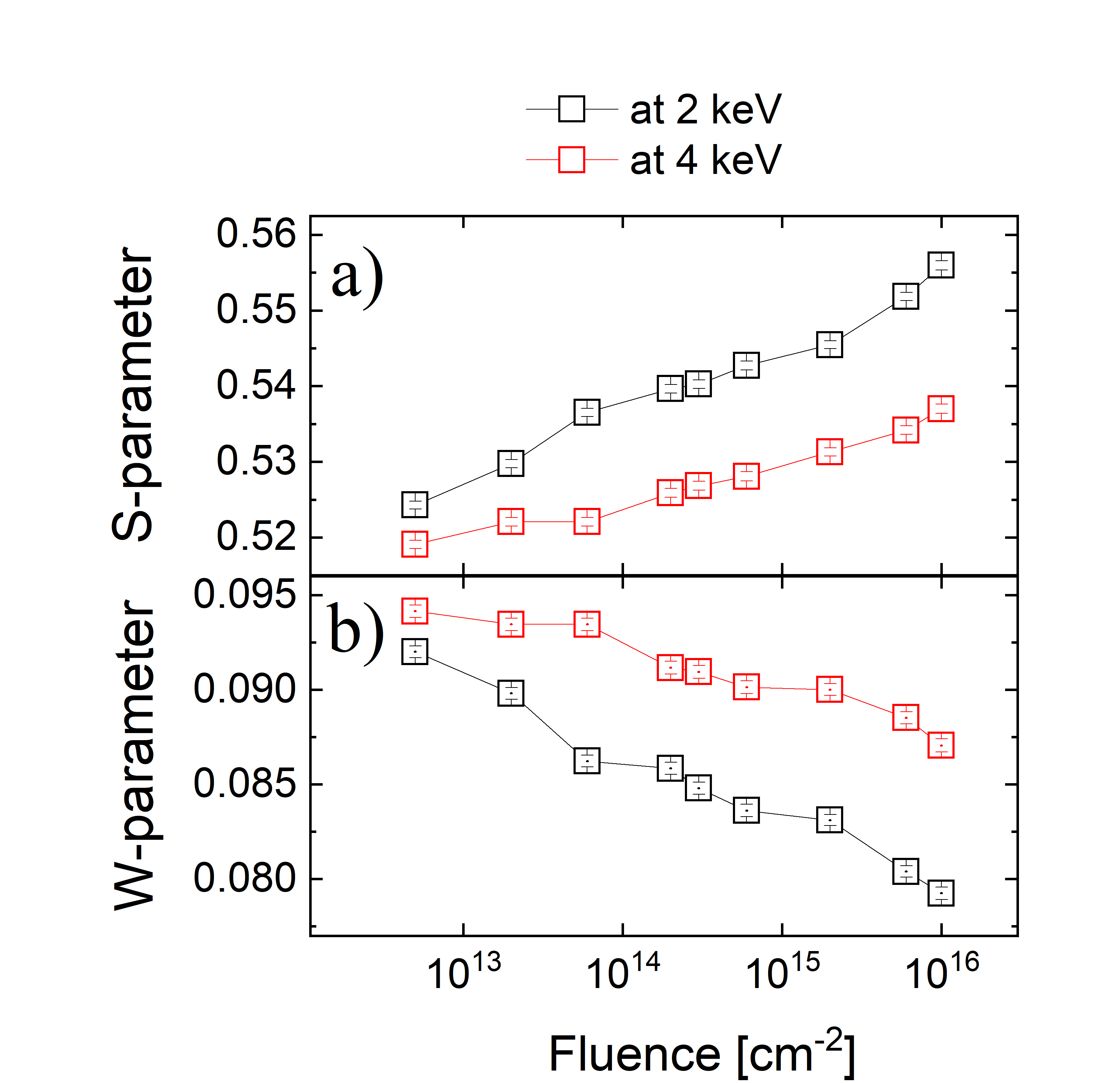}
        \caption{Dependence of S- and W-parameters on Yb implantation fluence for two selected positron energies of 2 and 4 keV, corresponding to mean depths of 15 and 50 nm, respectively. The results were obtained for ($\bar{2}$01)-oriented \bgao\ crystals. }
        \label{fig:dbvepas2}
\end{figure}

 Fig. \ref{fig:dbvepas2} presents the variations of S and W parameters as a function of Yb ion fluence used in ion implantation of \bgao, measured at two positron energies $E_p=2$ keV and $E_p=4$ keV. This approach allows for selective probing of defects in two depth regions of the sample. Specifically, $E_p=2$ keV probes the sub-surface layer, corresponding to the depth of approximately 15 nm, while $E_p=4$ keV probes deeper into the damaged zone, at approximately 50 nm. As can be seen, the low-momentum and high-momentum fractions are almost linear functions of ion fluence. This indicates that for the investigated range of Yb fluences, a similar defect type is expected. However, the fluence region between approximately \fluence{1}{14} and \fluence{1}{15} deviates from this linear trend, exhibiting a distinct plateau that indicates a structural rearrangement. 
 
The evolution of positron characteristics, disentangled from surface contribution, is also visible from the calculated S- and W-parameters for the sub-surface layer, which are presented in Fig. SM 3a and 3b in Supplementary Materials). 
Furthermore, Fig. SM 3d shows the calculated defect density, which increases several orders of magnitude in the \fluence{1}{14} - \fluence{1}{15} fluence region within the subsurface layer.
The defect density values were calculated for each Yb implantation fluence employing the formalism used previously for the same purpose \cite{bektas2025defect}.
The lower defect density in the deeper region may indicate the formation of extended defects like dislocation loops, which act as shallower traps for positrons. Furthermore, the numerical calculations indicate that around the fluence of \fluence{1}{15} the material reaches a defect saturation state in the subsurface layer. These observed thresholds of structural changes are in good agreement with RBS/C results, further confirming the depth-dependent nature of the radiation-induced damage.

\begin{figure}[h!]
        \centering
         \includegraphics[width=1\linewidth]{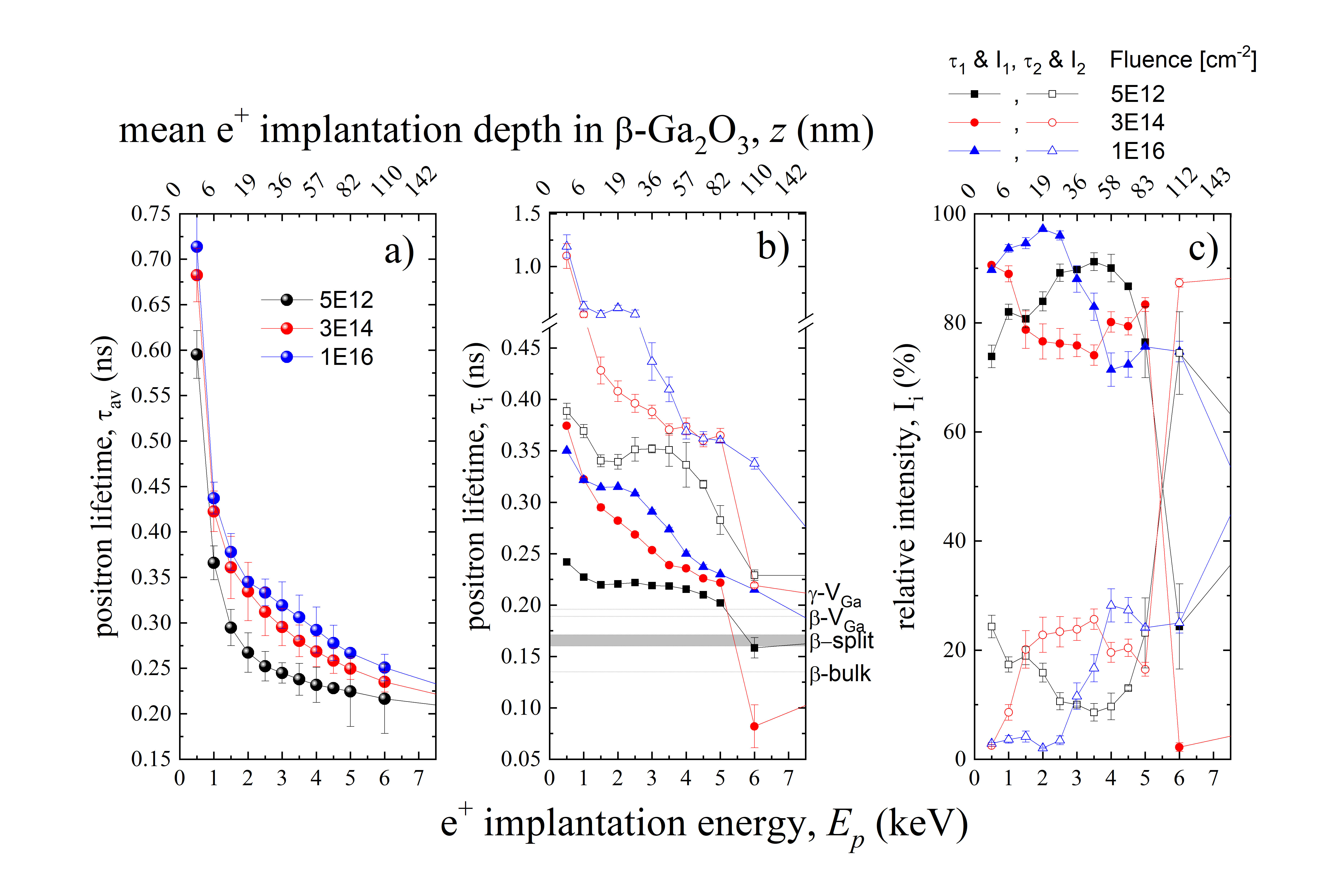}
        \caption{PALS depth profiles for selected fluences of Yb ions in ($\bar{2}$01)-oriented \bgao\ crystals: average positron lifetime $\tau_{av}$ (a), positron lifetime components $\tau_i$ (b) and their relative intensities $I_i$ (c). The calculated lifetimes for different defect structures in the $\beta$ and $\gamma$ phases are based on the work of \citet{bektas2025defect}.}
        \label{fig:pals1}
\end{figure}

Measurements of the positron annihilation lifetime (PALS, Positron Annihilation Lifetime Spectroscopy) provide information regarding both the types of defects and their concentration, as positron lifetime increases in open-volume defects due to lower electron density \cite{krause-rehberg2003positron}. PALS depth profiles were also obtained using varying e$^+$ implantation energy. As can be seen in Fig. \ref{fig:pals1}a, the average positron lifetime $\tau_{av}$ increases with fluence, which means that the average defect size increases with fluence as well. 
Moreover, $\tau_{av}$ decreases with depth, suggesting that defects in the deeper regions of the sample are smaller. 
To obtain more detailed information, the PALS spectra were decomposed into three lifetime components $\tau_i$, corresponding to different sizes of vacancy-related defects (see Fig. \ref{fig:pals1}b). Relative intensity $I_i$ (see Fig. \ref{fig:pals1}c) of a given lifetime component reflects the fraction of positrons trapped in a specific type of defect, allowing one to conclude about its concentration. Specifically, $\tau_1$ corresponds to small vacancy agglomerations larger than a single vacancy $V_{Ga}$, $\tau_2$ to large vacancy agglomerations, and $\tau_3$ to voids and surface contributions \cite{Privitera2026}. The concentration of voids and surface contributions is extremely low, and the measured intensity is close to the detection limit. For clarity, the contribution of voids is not shown in Fig. \ref{fig:pals1}, and it can be found in the Supplementary Materials (see Fig. SM 4). In contrast, the intensity of the $\tau_1$ component is the highest, indicating that this type of defect dominates in the Yb-implanted \bgao. It should also be noted that for the two higher fluences, the relative intensities vary with depth. That indicates the formation of two distinct regions within the implanted region, with different concentrations of defects. The data obtained at a fluence of \fluence{1}{16} clearly demonstrates that $I_1$ decreases at a specific depth, while $I_2$ increases, indicating the same that small vacancy clusters are dominant at the subsurface layer. This depth-dependent behavior is consistently observed in both RBS/C and PAS measurements.

\begin{figure}[h!]
        \centering
         \includegraphics[width=1\linewidth]{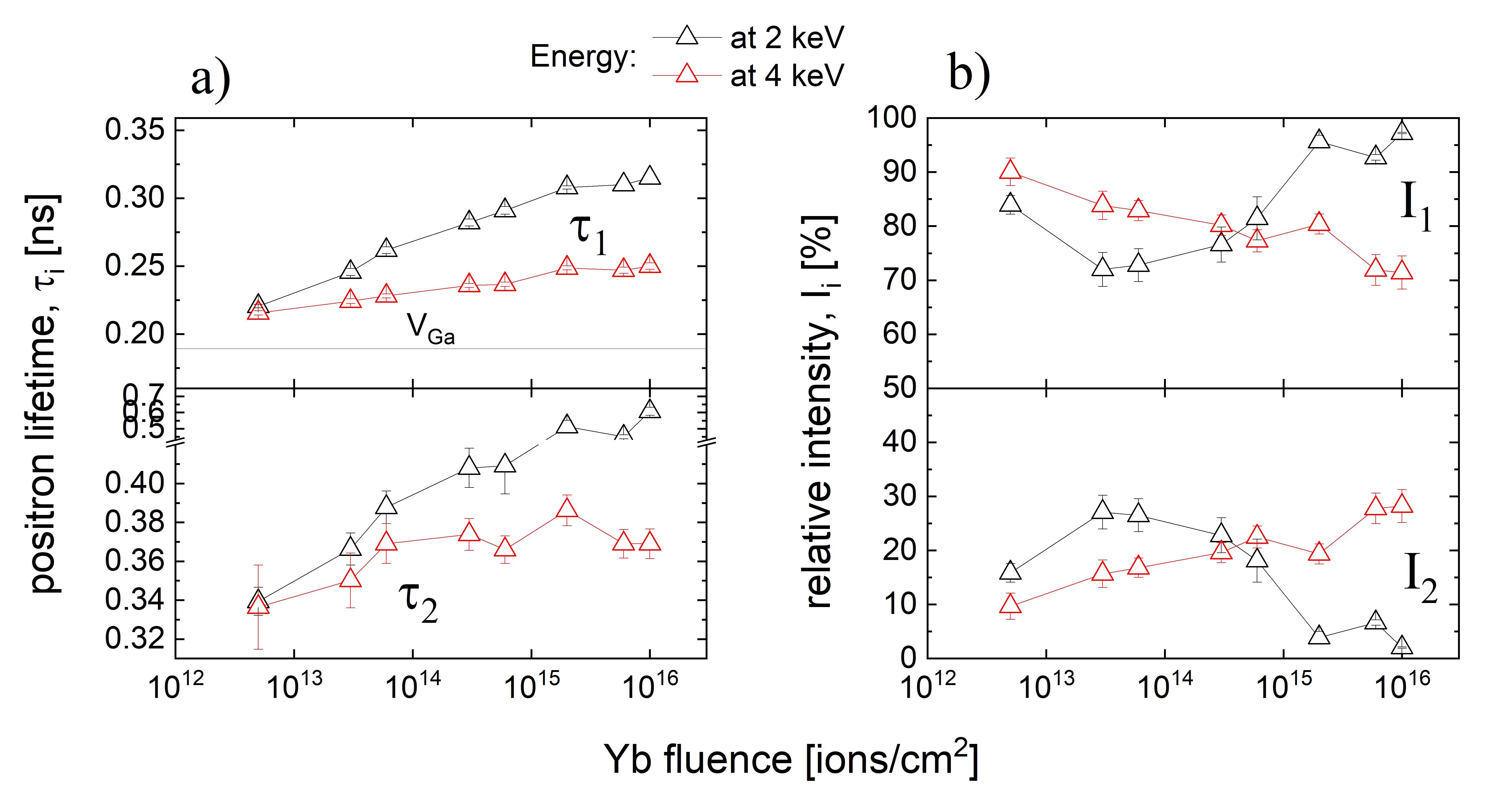}
        \caption{Dependence of $\tau_1$, $\tau_2$ and their relative intensities on Yb implantation fluence in ($\bar{2}$01)-oriented \bgao\ crystals, at two positron implantation energies of $E_p$=2 and 4 keV, corresponding to mean depths of about 15 and 50 nm, respectively.}
        \label{fig:pals2}
\end{figure}

To evaluate the depth-dependent distribution of the defect structure, two distinct positron implantation energies were selected, representing different regions of the damaged layer. Fig. \ref{fig:pals2} presents the two major lifetime components $\tau_1$, $\tau_2$, and their relative intensities $I_1$, $I_2$ at 2 and 4 keV, as a function of Yb ion fluence. Both $\tau_1$ and $\tau_2$ increase with increasing ion fluence. However, the slopes differ for the subsurface and deeper region, indicating different rates of defect accumulation within the implanted layer. 
Much more pronounced changes are observed in the behavior of the corresponding relative intensities. 
As shown in Fig. \ref{fig:pals2}b, for low fluences, the intensity values at both probing depths evolve comparably, and the differences between them remain minor.
However, above \fluence{3}{14}, a significant divergence emerges. 
Specifically, $I_1$ mostly increases in the subsurface region, whereas it decreases in the deeper region. 
The relative intensity $I_2$ exhibits the opposite trend. 
For the highest fluences, $I_1$ saturates at $\sim$95\% in the subsurface region, while $I_2$ reaches a very low level. Consequently, this indicates that $\tau_1$-related defects became the predominant positron trapping centers in the subsurface region. The PAS results presented here reveal significant structural disparities between the subsurface and deeper regions, which are consistent with observations obtained by RBS/C and other experimental methods.

\clearpage

\subsection{Identification of structural modifications in the implanted region}

\begin{figure}[h]
    \centering
    \includegraphics[width=0.9\linewidth]{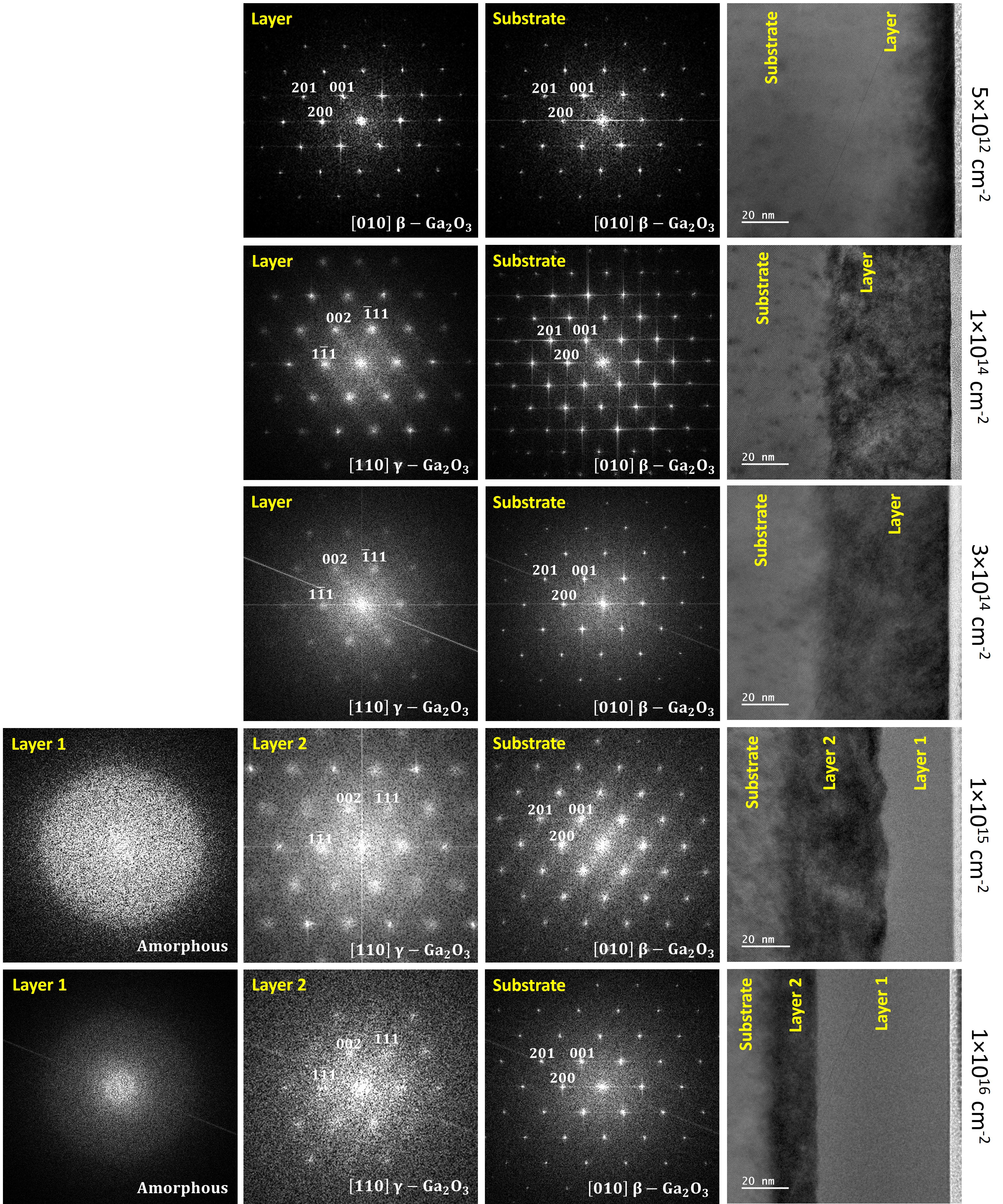}
    \caption{HRTEM images (on the right) and FFT patterns (on the left) of Yb-implanted ($\bar{2}$01)-oriented \bgao. }
    \label{fig:hrtem}
\end{figure}

\begin{figure}[h]
    \centering
    \includegraphics[width=0.5\linewidth]{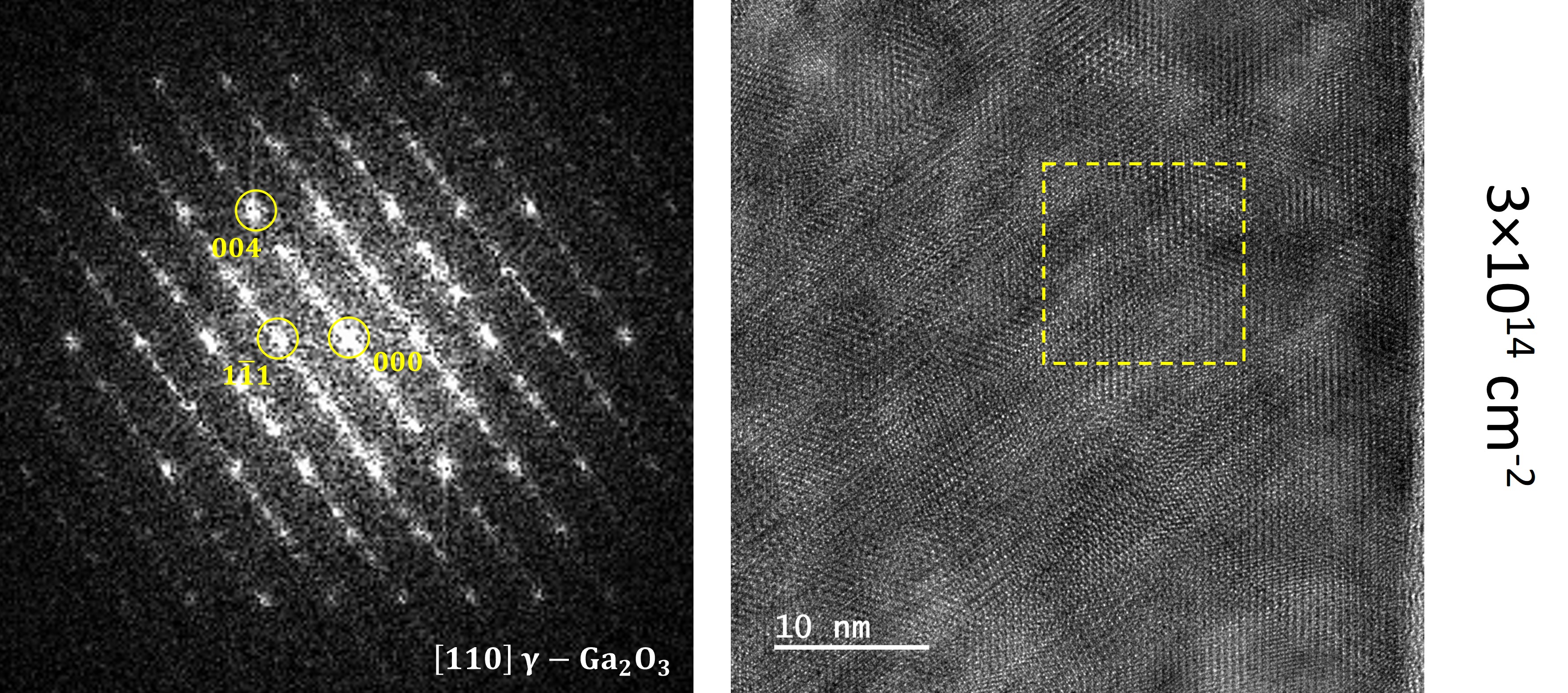}
    \caption{HRTEM image (on the right) and FFT pattern (on the left) of ($\bar{2}$01)-oriented \bgao\ implanted with \fluence{3}{14} Yb ions. }
    \label{fig:hrtem1}
\end{figure}

To identify structural modifications in the implanted region, HRTEM observations were performed for \bgao\ crystals implanted with different Yb fluences, which were selected based on the defect accumulation curve determined by RBS/C (Fig. \ref{fig:damage_accumulation}). The HRTEM images and the corresponding FFT patterns of ($\bar{2}$01)-oriented crystals are shown in Fig. \ref{fig:hrtem} and Fig. \ref{fig:hrtem1} (right and left panels, respectively). For comparison, the analysis was also performed for (010)-oriented crystal implanted with \fluence{1}{15}, shown in Fig. SM 2 in Supplementary Materials.

As can be seen in Fig. \ref{fig:hrtem}, for the low fluence of \fluence{5}{12} the implanted region is difficult to distinguish, which means that implantation with this fluence does not induce any significant structural changes in the material. This region is identified by HRTEM as \bgao. In contrast, for samples implanted with a fluence of \fluence{1}{14}, a modified layer became clearly visible in the HRTEM image, and the FFT analysis reveals that the entire implanted layer is the $\gamma$-phase of \gao. At a fluence of \fluence{3}{14}, the implanted layer remains in the $\gamma$-phase. However, the HRTEM image indicates the appearance of stacking faults in the subsurface region (see Fig. \ref{fig:hrtem1}). At the same fluence, a drop in the RBS/C defect accumulation curve (see Fig. \ref{fig:damage_accumulation}) is observed. TEM results clearly confirm that implantation with a higher fluence of \fluence{1}{15} leads to the formation of two distinct sublayers within the implanted region. No diffraction reflection can be found in the FFT pattern for the sublayer near the surface, while the deeper sublayer remains the \ggao\ layer. This observation is common to both ($\bar{2}$01)- and (010)-oriented crystals, indicating further phase transformation from \ggao\ to amorphous \gao, most likely driven by the accumulation of pre-existing stacking faults. At this fluence, both sublayers exhibit comparable thicknesses of approximately 30 nm, which agrees well with the RBS/C results \cite{ratajczak2024anisotropy}. Furthermore, based on the TEM analysis, it can be inferred that increasing the fluence further to \fluence{1}{16} causes the amorphous layer to grow at the expense of the $\gamma$ layer, which becomes thinner.

\clearpage

\subsection{Radiation-induced strain and other structural changes}

\begin{figure}[h]
    \centering
    \includegraphics[width=\linewidth]{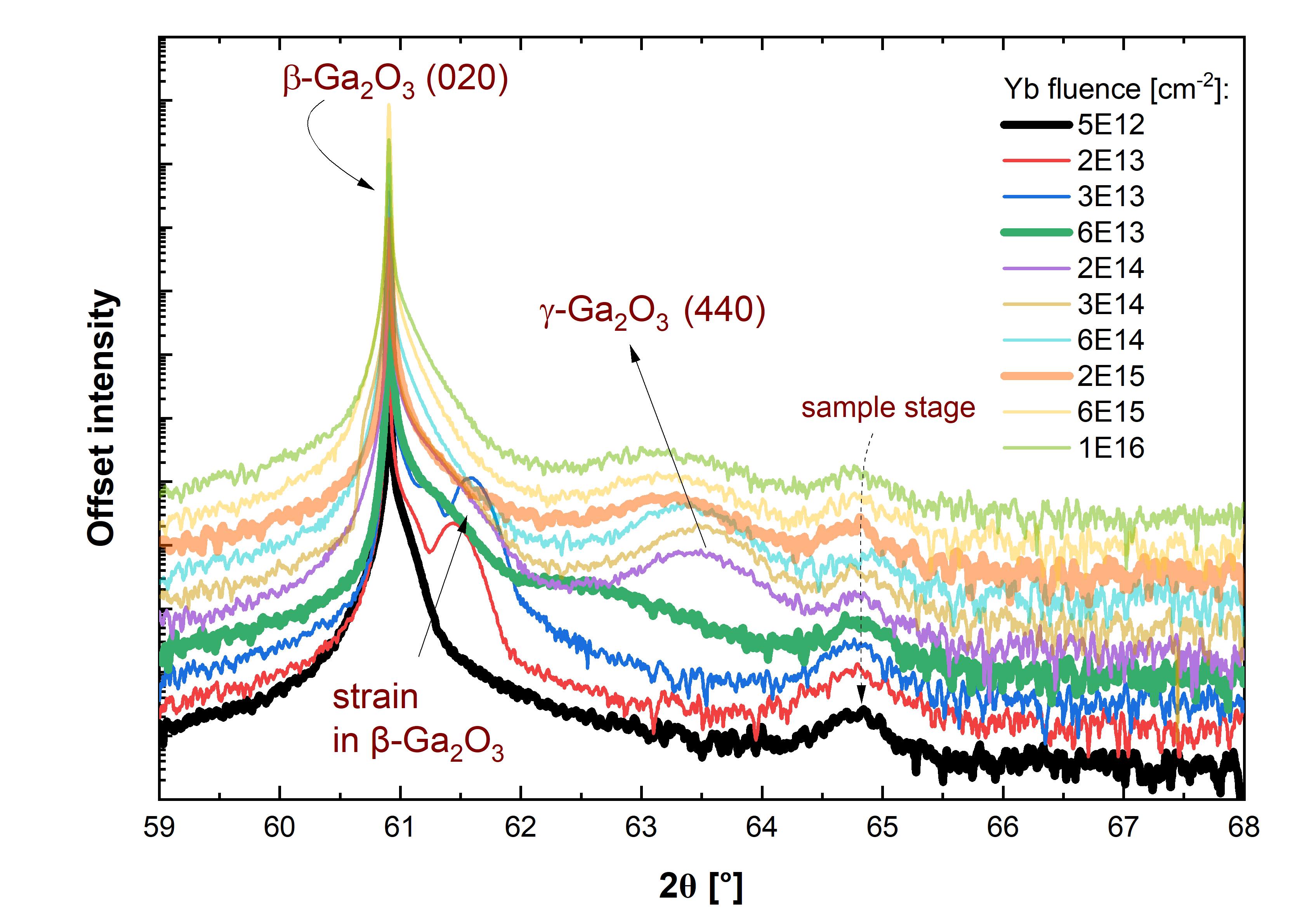}
    \caption{HRXRD scans of \bgao\ implanted with Yb ions of different fluences. The scans have been shifted along the vertical axis to make the differences between the results more visible. }
    \label{fig:xrd}
\end{figure}

HRXRD diffractograms of (010)-oriented \bgao\ substrate implanted with Yb ions at fluences ranging from \num{5}{12} to \fluence{1}{16} are shown in Fig. \ref{fig:xrd}. 

The main Bragg reflection at approximately 61\textdegree\ corresponds to the (020) reflection of the \bgao\ substrate. The sharp diffraction peak profile confirms the high crystalline quality of the initial material. 
The strain peak, which becomes increasingly pronounced at higher fluences, appears on the right side of the main Bragg reflection, indicating radiation-induced compressive strain. 
Interestingly, the HRXRD data indicate a minor strain even at the lowest fluence (\fluence{5}{12}) within the implanted region, despite the extremely low density of relative disorder observed by RBS/C. %Nie mamy porownania z probka nieimplantowana, wiec nie wiem czy warto o tym pisac
As the fluence increases up to \fluence{3}{13}, the strain peak shifts further towards higher 2$\theta$ angles and its intensity increases, revealing an increase not only in the magnitude of the strain but also in the volume of the material under strain.

For a fluence of \fluence{6}{13}, the strain signal disappears. Starting with the subsequent fluence \fluence{2}{14}, a clear signal centered around 63.5\textdegree\ becomes visible, corresponding to the (440) reflection of (110)-oriented \ggao. This is in good agreement with the literature on ion-implanted \bgao\ \cite{azarov2024optical}. As evidenced by earlier work \cite{sarwar2024defect} and HRTEM result presented here, the $\gamma$ phase is already present at a lower implantation fluence of \fluence{1}{14}, suggesting that strain relaxation initiates simultaneously with the formation of the $\gamma$ phase. Upon further irradiation up to \fluence{6}{14} the \ggao\ peak does not change significantly, which seems to indicate a stabilization of this intermediate phase. In fact, a decrease of the right shoulder on the \bgao\ signal can be observed. This suggests that a structural reorganization occurs within the \ggao\ crystal lattice, leading to strain relaxation and a subsequent reduction in the pressure exerted on the surrounding \bgao\ crystalline matrix. This strain relaxation may occur through the formation of stacking faults visible in TEM (Fig. \ref{fig:hrtem}).

Between fluences of \fluence{6}{14} and \fluence{2}{15}, the $\gamma$ peak profile undergoes a pronounced change, indicating a further structural evolution of this phase. The \ggao\ peak shifts slightly towards lower angles, and its intensity becomes significantly lower, which suggests a degradation of the material's crystalline quality.
For fluences \fluence{2}{15} the right shoulder on the \bgao\ remains low. However, at higher fluences, it increases again, indicating a renewed accumulation of lattice strain. The $\gamma$ diffraction peak remains detectable even for the highest fluence of \fluence{1}{16}, which is consistent with the presence of the residual \ggao\ layer observed in TEM. Thus, these changes between \fluence{6}{14} and \fluence{2}{15} can be associated with a transition from a crystalline to an amorphous phase, which is followed by the expansion of the amorphous phase at the expense of the $\gamma$ layer.

Although this detailed investigation of strain evolution as a function of ion fluence was conducted only for the (010)-oriented crystal, similar behavior might be expected for the (‑201) orientation, although differences cannot be excluded given the anisotropic and complex structure of \bgao.
%there is no reason to assume a different behavior for the ($\bar{2}$01) orientation.

\clearpage

\section{Discussion}
\label{disc}

\subsection{$\beta$ to $\gamma$ phase transformation}

Radiation-induced phase transition from $\beta-$ to \ggao\ was evidenced by several experimental methods. The presence of the $\gamma$ phase was directly identified in HRXRD diffractograms and HRTEM/FFT results for fluence of \fluence{1}{14} and higher. The reorganization of atoms leading to this phase transition is the reason for the rapid increase in relative disorder observed on the damage accumulation curve based on RBS/C. After the phase transition, the relative disorder reaches the random level, which is attributed to the misorientation of \ggao\ along the \bgao\ crystal axes, rather than a further increase in defect concentration. 
Based on HRXRD, it can be concluded that the observed phase transformation is driven by radiation-induced strain, which accumulates with increasing ion fluence. Once the transition is initiated, a subsequent strain relaxation is observed. This observation is consistent with the literature \cite{balog2024determination,huang2023atomic}. 

The measurements also allowed us to estimate the evolution of the phase transition. We assign the fluence of \fluence{6}{13}(0.44 dpa), where strain relaxation takes place, to the onset of the $\gamma$-phase formation. This value aligns closely with the value of 0.25 dpa predicted by MD simulations \cite{zhao2025crystallization}, where accumulated defects destabilize the monoclinic $\beta$ phase and trigger the structural transformation. Furthermore, the fluence of \fluence{1}{14} (0.74 dpa), at which the \ggao\ phase is clearly identified, corresponds well to the value of 0.65 dpa reported by \citet{zhao2025crystallization} for \ggao\ formation following Ni implantation into \bgao.

\subsection{$\gamma$ to amorphous \gao\ phase transformation} 

A further increase in the ion fluence leads to several interesting phenomena in \bgao\ implanted with Yb ions. Above a fluence of  \fluence{1}{14}, the density of vacancy-related defects starts to rapidly increase, while the relative disorder in the damage accumulation curve gradually decreases from 100\% up to approximately 85\%. Simultaneously, TEM analysis reveals the formation of stacking faults in the subsurface layer. Above \fluence{3}{14}, the relative disorder increases again, which is accompanied by the strain relaxation at the \bgao\ interface, visible in the HRXRD diffractograms. Therefore, we assign this drop in the damage accumulation curve to a partial alignment of Ga atoms along the main axis of \bgao\ due to atomic rearrangement underlying the subsequent structural transition, rather than to a sudden reduction in defect concentration. 

This second transformation appears to be completed at approximately \fluence{1}{15} (7 dpa), when the relative disorder in the defect accumulation curve reaches the random (fully disordered) level and the density of vacancy-related defects saturates at a high level. This is further confirmed by TEM, which clearly indicates that the accumulation of pre-existing stacking faults has culminated in the formation of an amorphous layer on the surface at this fluence. 

This outcome contrasts with previously published results, which show that amorphization does not occur even at dpa as high as 265 \cite{azarov2023universal}. 
This discrepancy cannot be attributed to the mass of the implanted ion, since in the mentioned work, no amorphization was observed even after Au implantation at 86 dpa. Therefore, our results challenge the concept of extreme radiation resistance of the $\gamma$ phase or suggest that its stability may depend on irradiation conditions, or potential chemical effects from RE dopants. The precise nature of this interaction requires further study. 
Previously, the amorphization of \bgao\ has also been suggested following Eu implantation \cite{lorenz2014doping,peres2017doping}. It was speculated that this high level of backscattering yield could be associated with Eu-induced defect stabilization or a reduced channeling efficiency due to the transition to the $\gamma$ phase \cite{azarov2023universal}. However, similarly to the RBS/C results for Yb-implanted \bgao\ at \fluence{1}{15}, the spectra for Eu-implanted \bgao\ at the same fluence exhibit two distinct peaks in the relative defect concentration \cite{peres2017doping}. This bimodal distribution strongly suggests the formation of an amorphous layer in the subsurface region, consistent with our observations for Yb. Additionally, the hypothesis of an amorphization threshold is further supported by the disappearance of the \ggao\ XRD signal after boron (B) implantation with a fluence corresponding to approximately 7 dpa \cite{nikolskaya2025structure}.

Our studies show that the further implantation with the fluences above 7 dpa causes a continuous expansion of the amorphous layer, which gradually replaces the crystalline \ggao. This is evidenced by the TEM image, as well as by the decreasing intensity of the \ggao\ diffraction peak in the intensity. Additionally, a slight shift of this peak toward lower angles in the XRD diffractogram is observed, also indicating increasing strain within the remaining crystalline regions as the amorphous phase expands.

\section{Conclusion}
\label{concl}

In this study, \bgao\ crystals were implanted with Yb ions with fluences ranging from \num{5}{12} to \fluence{1}{16} (0.04–74 dpa). 
They were subsequently investigated with a set of complementary experimental techniques, including RBS/C, PAS, HRTEM, and HRXRD, providing a comprehensive description of the radiation-induced damage accumulation process in Yb-implanted \bgao. 
The results reveal a complex process of defect evolution that proceeds through four stages, and involves two sequential phase transformations: from the initial $\beta$ phase to the $\gamma$ phase, followed by the amorphization of the $\gamma$ layer. Notably, these transitions occur independently of the crystal orientation, and are driven primarily by radiation-induced strain.

The observed amorphization of \bgao\ via \ggao\ stands in contrast to several previous studies in which the \ggao\ crystalline structure remained intact even under irradiation with 265 dpa. The reason for this discrepancy may be attributed to specific ion-solid interactions and will be the subject of further research.

\section*{Acknowledgments}
The research was carried out within the NCN project UMO-2022/45/B/\\ST5/02810. 

D.K. and M.A.S. were supported by the European Union Horizon 2020 research and innovation program under Grant Agreement No 857470 and from the European Regional Development Fund under the program of the Foundation for Polish Science International Research Agenda PLUS, grant No MAB PLUS/2018/8, and the initiative of the Ministry of Science and Higher Education 'Support for the activities of Centers of Excellence established in Poland under the Horizon 2020 program' under agreement No MEiN/2023/DIR/3795.

Parts of this research were carried out at ELBE at the Helmholtz-Zentrum Dresden – Rossendorf e. V., a member of the Helmholtz Association. We would like to thank the facility staff for assistance. This work was partially supported by the Initiative and Networking Fund of the Helmholtz Association (FKZ VH-VI-442 Memriox), and the Helmholtz Energy Materials Characterization Platform (03ET7015).

\bibliographystyle{unsrt} 
\bibliography{bibliography}



\section*{Supplementary material}

**Supplementary Materials**

# From phase transformation to amorphization: damage accumulation in Yb-implanted $\beta$-$Ga_2O_3$

J. Matulewicz, R. Ratajczak, E. Grzanka, M. O. Liedke, D. Kalita, E. Hirschmann, A. Wagner, M. Grabowski, M. A. Stróżyk, C. Mieszczyński, P. Jóźwik, U. Kentsch, R. Heller, F. Garrido

## 1 SRIM simulations

SRIM simulations [1] were performed to obtain the dpa (displacement per atom) value for 150 keV Yb ions implanted in $\beta$-$Ga_2O_3$. Fig. 1 presents the distribution of Yb ions, fitted to a Gaussian, with a maximum value at 31 nm, and the dpa calculated from the atomic displacement distribution (including vacancies and atomic replacements from the VACANCY.txt and NOVAC.txt files). The simulations were done using the default values of Ga and O threshold displacement energies ($E_d$) given by the SRIM software, 25 and 28 eV, respectively.

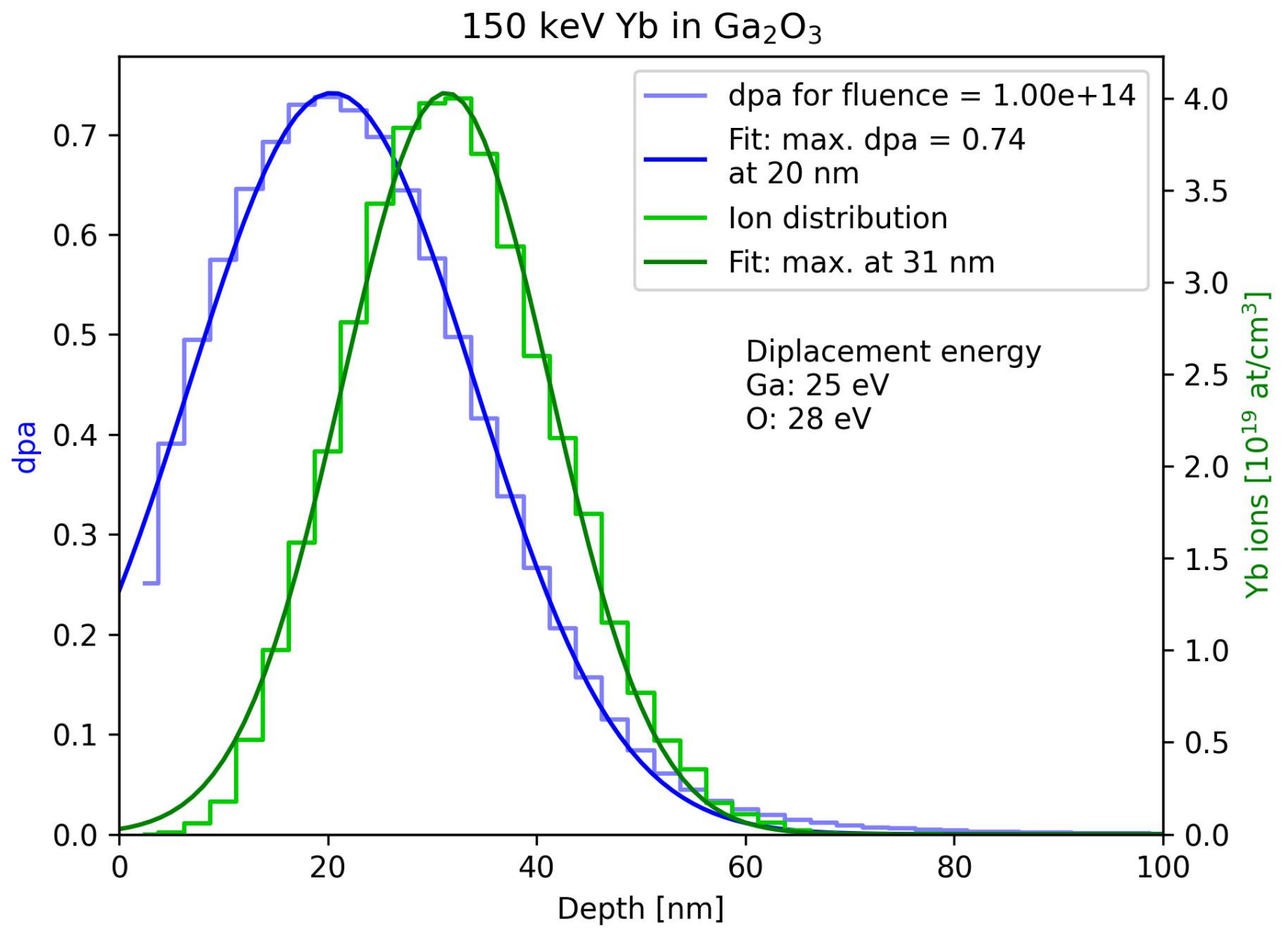


Figure SM 1: Yb ion distribution and dpa, obtained using SRIM [1].

# 2 HRTEM

For comparison, HRTEM imaging was done also for (010)-oriented $\beta$-$Ga_2O_3$ crystal implanted with $1 \times 10^{15}$ $cm^{-2}$. Just like for ($\bar{2}$01)-oriented $\beta$-$Ga_2O_3$, the HRTEM image in Fig. 2 shows three layers. The corresponding FFT patterns confirm the presence of an amorphous layer on the surface and a $\gamma$-$Ga_2O_3$ layer.

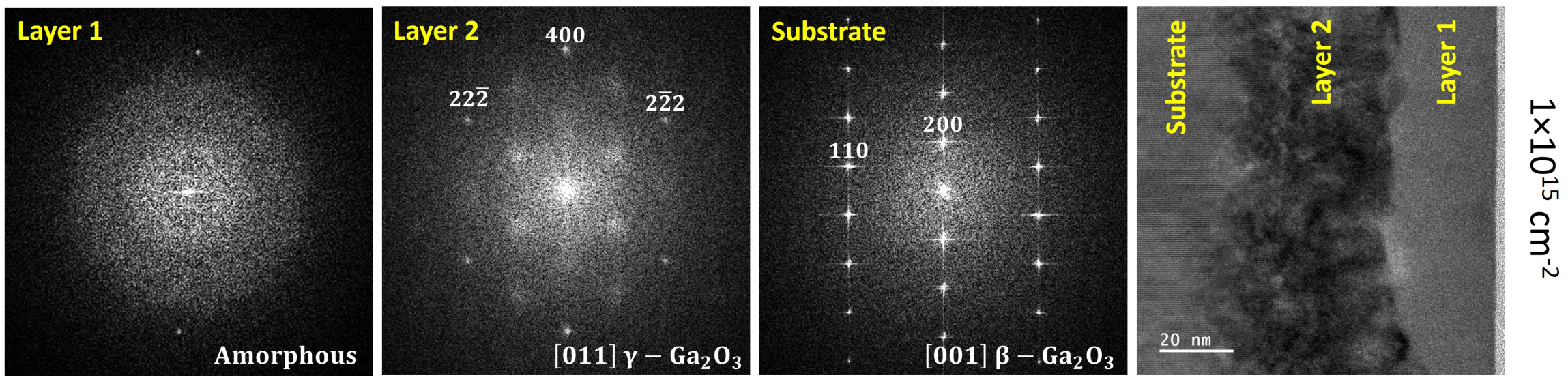


Figure SM 2: HRTEM image and FFT patterns of (010)-oriented $\beta$-$Ga_2O_3$ implanted with Yb with fluence of $1 \times 10^{15}$ $cm^{-2}$.

# 3 PAS

## 3.1 Calculation of the positron diffusion length using VEPfit

For the analysis of positron diffusion length, $L_+$, which scales inversely to defect concentration, the VEPFit code was used [2, 3]. It allows to fit $S(E_p)$ and $W(E_p)$ curves for multilayered systems, and to acquire thickness, $L_+$, and specific S- and W-parameters for each layer within a stack. The calculated defect densities of the sub-surface layer $c_{v,sub}$ and bulk $c_{v,bulk}$, based on respective $L_+$ and corresponding defected layer thickness $d$, are presented in Table SM 1 and in Fig. SM 3.

The evolution of positron characteristics, disentangled from surface contributions, is visible from the calculated S- and W-parameters for the sub-surface layer (see Fig. 3a and b). The calculated S-parameter is quite small, exhibiting a subtle rising trend until the fluence of $6 \times 10^{13}$ $cm^{-2}$, followed by a step-like increase for $2 \times 10^{14}$ and $3 \times 10^{14}$ $cm^{-2}$, and a further rise for larger fluences. The calculated W-parameter has a general opposite trend, besides the abrupt change for the fluence of $3 \times 10^{14}$ $cm^{-2}$. The rise of the calculated S and decrease of W coincide very well with the damage accumulation curve. This behavior implies the high non-uniform defect distributions, where small vacancy-related defects are localized in the subsurface region. The calculated thickness $d$ of that sub-surface layer increases nearly linearly to about $d = 270$ nm until the fluence of $2 \times 10^{13}$ $cm^{-2}$ and subsequently decreases to about $d = 5.7$ nm at $3 \times 10^{14}$ $cm^{-2}$, remaining in the range of about 13-21 nm for larger fluences (see Fig. 3c).

## 3.2 Calculation of defect concentration based on positron diffusion length

Assuming that the films mostly contain one major defects type, the vacancy concentration $c_V$ can by calculated as follows [4]:

$$c_V = \frac{N}{\nu_V \tau_B} \left( \frac{{L_{+,B}}^2}{{L_+}^2} - 1 \right),$$

where $N = 4.7710^{22}$ at×$cm^{-3}$ and $N = 5.7210^{22}$ at×$cm^{-3}$ are atomic densities of $\beta$-$Ga_2O_3$ and $\gamma$-$Ga_2O_3$, $\nu_V \approx 1 \times 10^{15}$ $s^{-1}$ is a specific positron trapping rate (trapping coefficient) [5], $\tau_B \approx 135$ ps and $\tau_B \approx 173$ ps are

Table SM 1: S- and W-parameters of the subsurface region of thickness $d$, effective diffusion lengths and corresponding defect densities for all samples, calculated using VEPFit code [2].

| Yb fluence [cm$^{-2}$] | $S_{sub-surf.}$ | $S_{err}$ | $W_{sub-surf.}$ | $W_{err}$ | $d$ [nm] | $d_{err}$ | $L_+$ [nm] | $L_{+,bulk}$ [nm] | $c_{v,sub}$ [cm$^{-3}$] | $c_{v,bulk}$ [cm$^{-3}$] |
|---|---|---|---|---|---|---|---|---|---|---|
| 1E12 | 0.5126 | 2E-4 | 0.0972 | 1E-4 | 0 | – | 29±0.9 | 69 | – | 3.9 |
| 5E12 | 0.5158 | 2E-4 | 0.0963 | 1E-4 | 175 | 8 | 45±0.7 | 69 | 38 | 3.9 |
| 2E13 | 0.5136 | 2E-4 | 0.0975 | 1E-4 | 270 | 17 | 38±0.6 | 69 | 14 | 3.9 |
| 6E13 | 0.5164 | 5E-4 | 0.0959 | 2E-4 | 92 | 19 | 16.8±0.5 | 65±2 | 21 | 3.9 |
| 2E14 | 0.5319 | 4E-4 | 0.09 | 2E-4 | 38 | 4 | 3.8±0.3 | 58±1 | 122 | 4.8 |
| 3E14 | 0.5322 | 0.0017 | 0.0968 | 1E-4 | 5.7 | 0.5 | 1.7±0.2 | 56±1 | 2443 | 7 |
| 6E14 | 0.5453 | 2E-4 | 0.0823 | 1E-4 | 13.5 | 0.7 | 0.9±0.1 | 57±1 | 12222 | 7.7 |
| 2E15 | 0.5471 | 2E-4 | 0.0818 | 1E-4 | 18 | 0.8 | 1.3±0.1 | 48±1 | 43615 | 7.3 |
| 1E16 | 0.5585 | 2E-4 | 0.0775 | 1E-4 | 20.8 | 0.7 | 1.2±0.3 | 52±1 | 24532 | 9.5 |

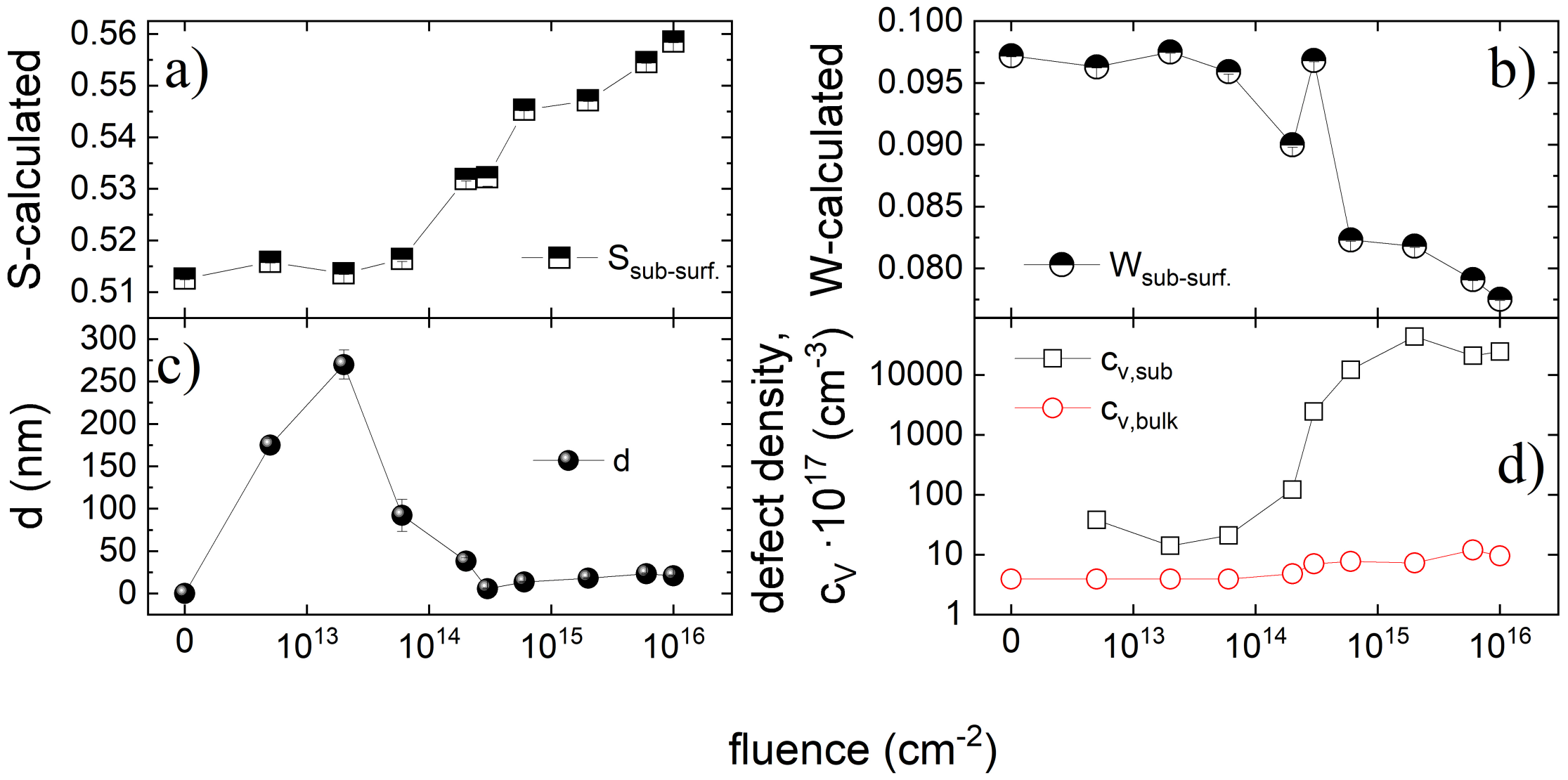


Figure SM 3: The calculated S and W-parameters, damaged sub-surface layer thickness $d$, and respective calculated defect densities as a function of Yb fluence

bulk lifetimes for $\beta$-$Ga_2O_3$ and $\gamma$-$Ga_2O_3$ and $L_{+,B} = 100$ nm is a diffusion length in our best reference sample.

## 3.3 Methods

### 3.3.1 Doppler broadening variable energy positron annihilation spectroscopy (DB-VEPAS)

Doppler broadening variable energy positron annihilation spectroscopy (DB-VEPAS) measurements have been conducted at the apparatus for in-situ defect analysis (AIDA) [6] of the slow positron beamline (SPONSOR) [7]. Positrons have been accelerated and monoenergetically implanted into samples in the energy $E_p$ range 0.05 keV - 35 keV, which allows for depth profiling. The mean positron implantation depth was approximated using

a simple material density ($\rho$) dependent formula [8]:

$$\langle z \rangle \,[\text{nm}] = \frac{36}{\rho[\text{g} \times \text{cm}^{-3}]} E_p[\text{keV}]^{1.62}.$$

$\langle z \rangle$ approximates the depth and cannot be treated as an absolute measure because it does not account for positron diffusion. It gives the best estimation for materials with a large defect concentration, hence low positron diffusion length. Positrons implanted into a solid lose their kinetic energy due to thermalization, and after short diffusion, they annihilate in delocalized lattice sites, or localize in vacancy-like defects and interfaces, emitting usually two anti-collinear 511 keV gamma photons once they meet electrons. Since at the annihilation site thermalized positrons have very small momentum compared to the electrons, a broadening of the 511 keV line is observed mostly due to the momentum of the electrons, which is measured with one or two high-purity Ge detectors (energy resolution of 1.09 ± 0.01 keV for single detector configuration, and 0.78 ± 0.02 keV at 511 keV for double). This broadening is characterized by two distinct parameters S and W, defined as a fraction of the annihilation line in the middle (511 ± 0.93 keV) and outer regions (504.04-508.68 keV and 513.32-517.96 keV), respectively. The total area below the curve, which is used for the normalization of both parameters, is 511 ± 16.24 keV. The S-parameter is a fraction of positrons annihilating with low-momentum valence electrons, and represents vacancy- type defects and their concentration. The W-parameter approximates the overlap of the positron wavefunction with the high-momentum core electrons.

### 3.3.2 Variable energy positron annihilation lifetime spectroscopy (VEPALS)

Variable energy positron annihilation lifetime spectroscopy (VEPALS) measurements were conducted at the Mono-energetic Positron Source (MePS) beamline at HZDR, Germany [9]. A $CeBr_3$ scintillator coupled to a Hamamatsu R13089-100 photomultiplier tube (PMT) was used for the gamma photons detection. The signals were processed by the Teledyne SP Devices ADQ14DC-2X digitizer (14 bit vertical resolution and 2GS/s horizontal resolution) [10]. The overall time resolution of the measurement system is $\approx 0.250$ ns and all spectra contained at least $1 \times 10^7$ counts. A typical lifetime spectrum $N(t)$, the absolute value of the time derivative of the positron decay spectrum, is described by the formula

$$N(t) = R(t) \times \sum_{i=1}^{k+1} \frac{I_i}{\tau_i} e^{\frac{-t}{\tau_i}} + \text{background},$$

where $k$ is the number of different defect types contributing to the positron trapping, which are related to $k+1$ components in the spectra with the individual lifetimes $\tau_i$ and intensities $I_i$ ($\sum I_i = 1$) [11]. The instrument resolution function $R(t)$ is a sum of two Gaussian functions with distinct intensities and relative shifts, both depending on the positron implantation energy, $E_p$. It was determined by the measurement and analysis of a reference sample, i.e., Yttria-stabilized zirconia (YSZ), which exhibited a single well-known lifetime component. The background was negligible, hence fixed to zero.

All the spectra were deconvoluted using a non-linear least-squares fitting method, minimized by the Levenberg-Marquardt algorithm, employed within the fitting software package PALSfit [12] into two major lifetime components, which directly evidence localized annihilation at two different defect types (sizes; $\tau_1$ and $\tau_2$). Their relative intensities scale typically with the concentration of each defect type. In general, positron lifetime increases with defect size and open volume size [13]. The positron lifetime and its intensity have been probed as a function of positron implantation energy $E_p$, which was recalculated to the mean implantation depth $\langle z \rangle$. The average positron lifetime $\tau_{av}$ is defined as $\tau_{av} = \sum_i \tau_i \times I_i$.

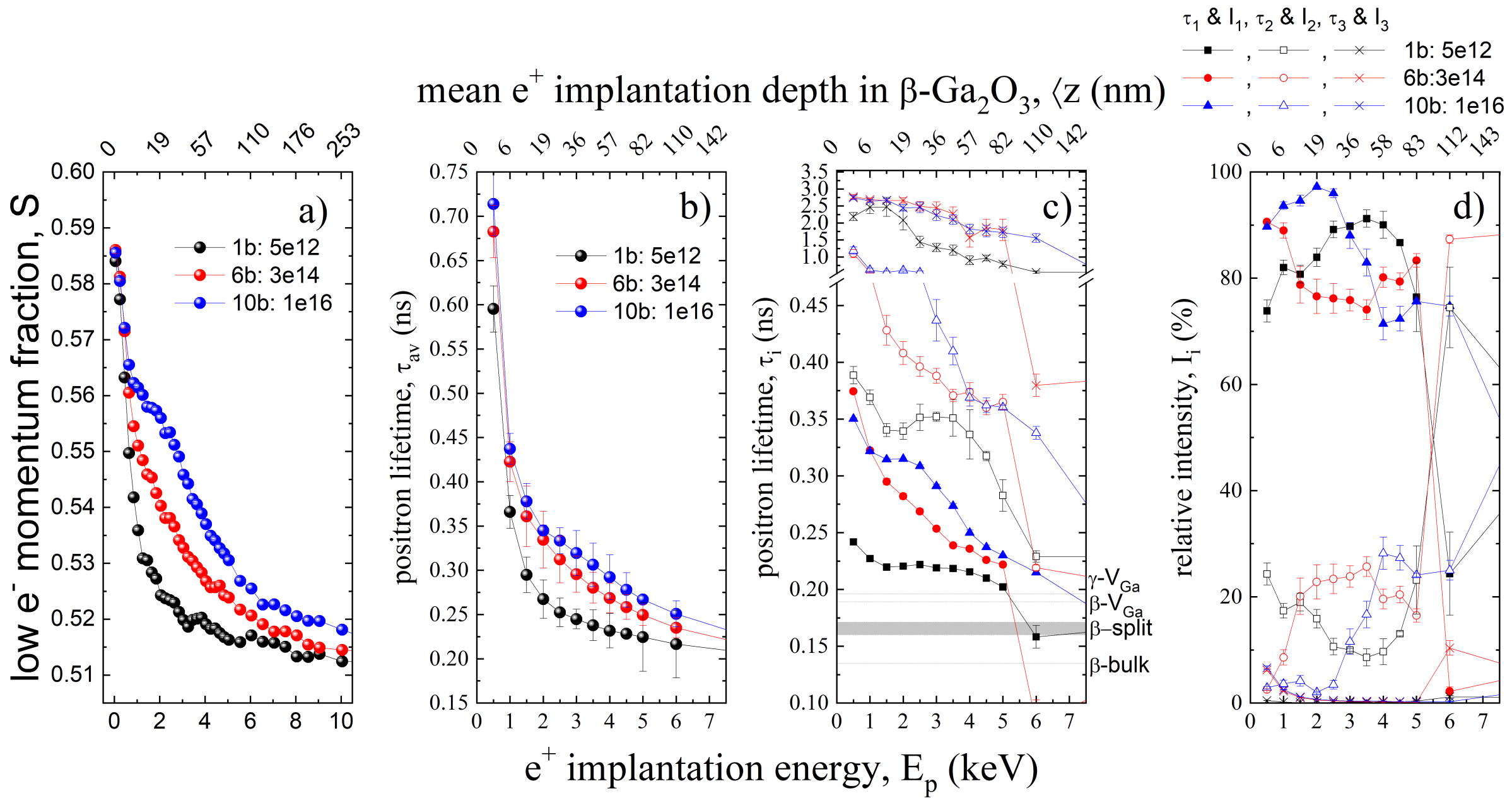


Figure SM 4: PALS depth profiles for selected Yb ion fluences in $(\bar{2}01)$-oriented $\beta$-$Ga_2O_3$ crystals: average positron lifetime $\tau_{av}$ (a), positron lifetime components $\tau_i$ (b) and their relative intensities $I_i$ (c). The calculated lifetimes for different defect structures in the $\beta$ and $\gamma$ phases are based on the work of Bektas et al. [14].



\end{document}